\documentclass{article}
\pdfoutput=1
\usepackage{graphicx}
\usepackage{placeins}
\usepackage{hyperref}
\hypersetup{colorlinks=true,linkcolor=blue, filecolor=blue, urlcolor=blue, citecolor=blue, pdfauthor={J.C.Smallwood}, pdftitle={Test-beam demonstration of a TORCH prototype module}}

\usepackage[symbol]{footmisc}

\usepackage{geometry}
\usepackage{multirow}
\usepackage{amsmath}
\usepackage{comment}
\usepackage[backend=biber, style=phys, natbib=true, maxnames=4]{biblatex} 
\begin{document}
\textit{Prepared for submission to IoP Conference Series}\\
International Conference on Technology and Instrumentation in Particle Physics\\
24 - 28 May 2021 \\
TRIUMF, Vancouver \\

\bigskip

\textbf{\large{Test-beam demonstration of a TORCH prototype module}}

\bigskip

\rule{\linewidth}{1mm}

\bigskip

\textbf{J.~C.~Smallwood$^{a,}$\footnote[1]{Corresponding author \\ \hspace{10mm} \textit{Email address:} jennifer.smallwood@cern.ch },
S.~Bhasin$^{b,c}$,
T.~Blake$^f$,
N.~H.~Brook$^b$,
M.~F.~Cicala$^f$, 
T.~Conneely$^g$, \\
D.~Cussans$^c$,
M.~W.~U.~van~Dijk$^d$,
R.~Forty$^d$,
C.~Frei$^d$, 
E.~P.~M.~Gabriel$^e$, 
R.~Gao$^a$, \\
T.~Gershon$^f$,
T.~Gys$^d$,
T.~Hadavizadeh$^a$, 
T.~H.~Hancock$^a$, 
N.~Harnew$^a$, 
M.~Kreps$^f$, \\
J.~Milnes$^g$,
D.~Piedigrossi$^d$,
J.~Rademacker$^c$}

\bigskip

\textit{$^a$Denys Wilkinson Laboratory, University of Oxford, Keble Road, Oxford, OX1 3RH, United Kingdom} \\
\textit{$^b$University of Bath, Claverton Down, Bath, BA2 7AY, United Kingdom} \\
\textit{$^c$H.H. Wills Physics Laboratory, University of Bristol, Tyndall Avenue, Bristol, BS8 1TL, United Kingdom} \\
\textit{$^d$European Organisation for Nuclear Research (CERN), CH-1211 Geneva 23, Switzerland} \\
\textit{$^e$School of Physics and Astronomy, University of Edinburgh, James Clerk Maxwell Building, Edinburgh, EH9 3FD, United Kingdom} \\
\textit{$^f$Department of Physics, University of Warwick, Coventry, CV4 7AL, United Kingdom} \\
\textit{$^g$Photek Ltd., 26 Castleham Road, St Leonards on Sea, East Sussex, TN38 9NS, United Kingdom} \\

\medskip

\begin{abstract}

The TORCH time-of-flight detector is designed to provide a 15 ps timing resolution for charged particles, resulting in $\pi$/K particle identification up to 10 GeV/c momentum over a 10 m flight path. Cherenkov photons, produced in a quartz plate of 10 mm thickness, are focused onto an array of micro-channel plate photomultipliers (MCP-PMTs) which measure the photon arrival times and spatial positions. A half-scale ($660 \times 1250 \times 10$ mm$^3$) TORCH demonstrator module has been tested in an 8 GeV/c mixed proton-pion beam at CERN. Customised square MCP-PMTs of active area $53 \times 53$ mm$^2$ and granularity $64 \times 64$ pixels have been employed, which have been developed in collaboration with an industrial partner. The single-photon timing performance and photon yields have been measured as a function of beam position in the radiator, giving measurements which are consistent with expectations. The expected performance of TORCH for high luminosity running of the LHCb Upgrade II has been simulated.
\end{abstract}

\newpage
\setlength{\parindent}{2em}
\section{Introduction}

\begin{figure}[h]
\begin{center}
\begin{minipage}{2.0in}
\includegraphics[width=1\linewidth]{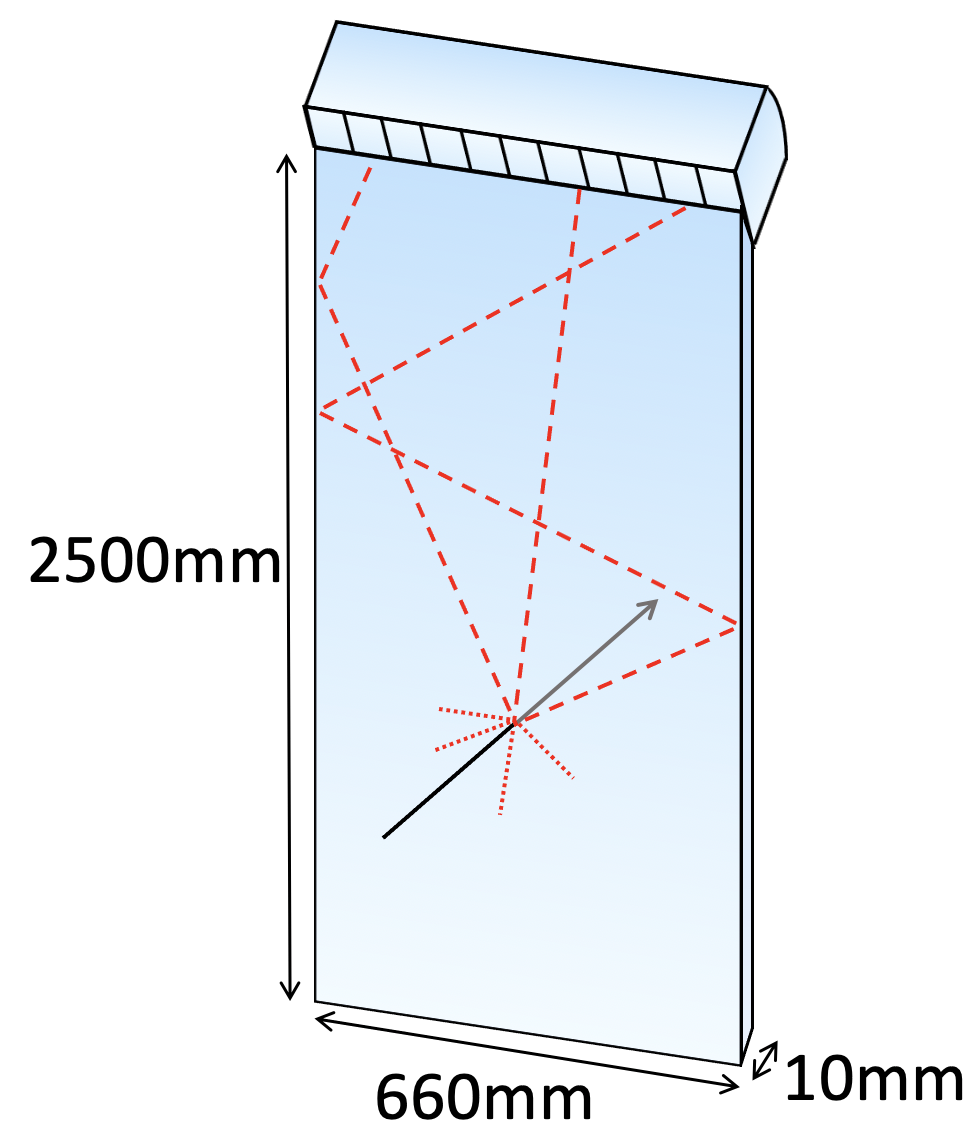} 
\end{minipage}\hspace{2pc}%
\begin{minipage}{1.7in}
\includegraphics[width=1\linewidth]{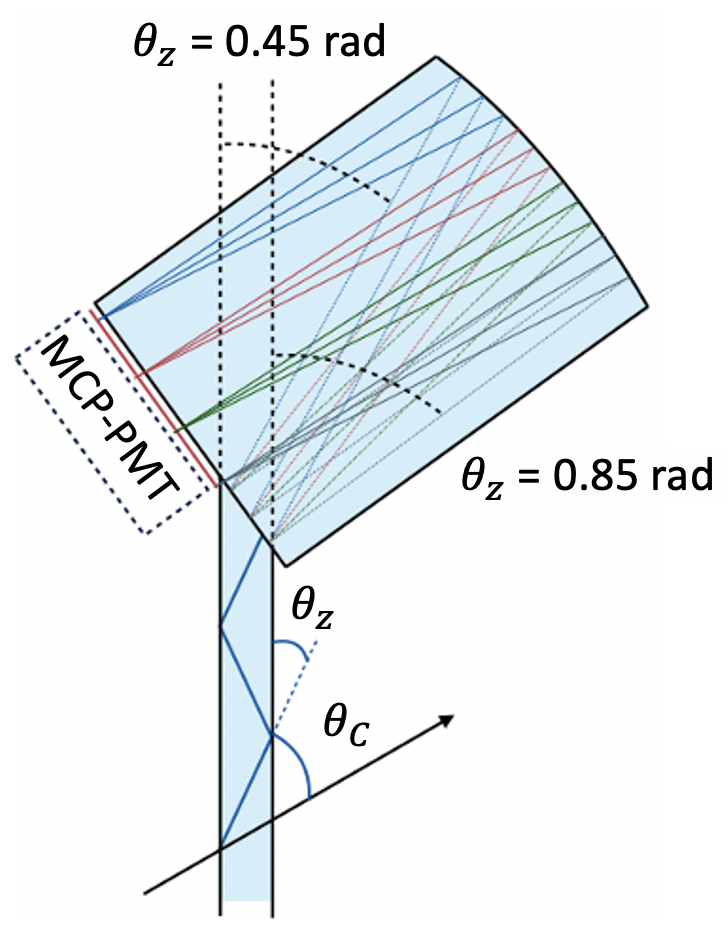}
\end{minipage}
\caption{\label{Torch} Schematics demonstrating the principle of TORCH. (Left) Charged particles traversing the radiator plate produce Cherenkov photons which travel to the focusing optics by total internal reflection. (Right) The focusing optics map the photon angle onto position on the detector plane such that the Cherenkov angle of emission can be determined.}
\end{center}
\end{figure}

TORCH is a large-area time-of-flight (ToF) detector proposed for the LHCb experiment \cite{TheLHCbCollaboration2011}. It will provide particle identification (PID) of charged hadrons between approximately $2-20$~GeV/c momentum. 
A basic TORCH module consists of a large quartz plate with focusing optics and photon detectors positioned at one end, as shown in Fig.~\ref{Torch}. Charged particles will emit Cherenkov photons when traversing the plate. Accepted photons will travel via total internal reflection to a focusing block where they are reflected by a mirrored surface onto the detector plane. There the positions of the photons are measured, from which the Cherenkov angle and path length within the quartz can be determined for each photon. The time of arrival is then used to correct for chromatic dispersion effects and the time of propagation calculated \cite{Charles2011} \cite{Brook2018}. 

Micro-channel plate photomultiplier tubes (MCP-PMTs) are used to detect the Cherenkov photons. These have been developed by industrial partner Photek UK (Ltd) to TORCH specifications \cite{Conneely2015}. Each MCP-PMT has a square $53\times53$~mm$^2$ active area on a pitch of $60\times60$~mm$^2$, with a granularity of $64\times64$ pixels. The requirement on the resolution in the horizontal (non-focusing) direction is such that channels can be grouped electronically to a granularity of $64\times8$. Charge sharing across pixels in the vertical (focusing) direction is designed to give an effective granularity of $128\times8$ which improves the spatial and timing precision of the detected photons. 
The MCP-PMTs have been manufactured with an atomic layer deposition (ALD) coating to withstand an integrated charge accumulation of at least 5 Ccm$^{-2}$. 

The detectors are read out with custom electronics \cite{Gao2016} that use the NINO \cite{Anghinolfi2004} and HPTDC \cite{Moreira2000} chipsets.
The NINO provides time-over-threshold information and corrections are made to account for the pulse-height-dependent signal shape (time-walk). Corrections are also made for non-linearities in the HPTDC binning. 

TORCH has been proposed for the Upgrade II of the LHCb experiment. Eighteen $660\times 2500\times 10$ mm$^3$ modules will be positioned approximately 9.5 m from the interaction region, each equipped with 11 MCP-PMTs. 
The time of flight difference between pions and kaons at 10 GeV/c over this distance is $\sim 35$ ps, requiring a time-resolution per track of at least 15 ps for clean separation. 
This can be achieved for 30 detected photons per track with a single-photon timing resolution of 70 ps, assuming a $\sim \frac{1}{\sqrt{N}}$ dependence. 

\section{The Test Beam Campaign}

A half-height, full-width ($660\times 1250\times 10$ mm$^3$) prototype module (called ``Proto-TORCH'') was tested at the CERN PS East Hall T9 facility in October 2018. Two MCP-PMTs (referred to as MCP A and MCP B) were instrumented, positioned on the left-hand edge when viewing Proto-TORCH from its front face. 

The test beam infrastructure makes use of two time reference stations constructed from thin borosilicate bars in which Cherenkov light is detected by a single channel MCP-PMT. These were positioned 10 m upstream and 1 m downstream of Proto-TORCH. In addition, a pair of Cherenkov counters provided an independent source of PID for the beam, which was composed of approximately 46:54 pion:protons at a momentum of 8 GeV/c. A pixel telescope from EUDET/AIDA \cite{Rubinskiy2014} was used to measure the beam profile.

Data were taken with the beam incident at specific positions on the plate, with the aim of measuring the single-photon time resolution and the photon yields. These positions are numbered such that positions 1, 3, 4 and 5 are positioned 5mm inset into the plate on the side closest to the MCP-PMTs, with vertical distance of 175 mm, 489 mm, 802 mm and 1115 mm from the top of the plate respectively. Position 6 is located in the centre of the plate horizontally, and at a vertical distance of 1115 mm.

\begin{figure}[h]
\begin{center}
\includegraphics[width=0.55\linewidth]{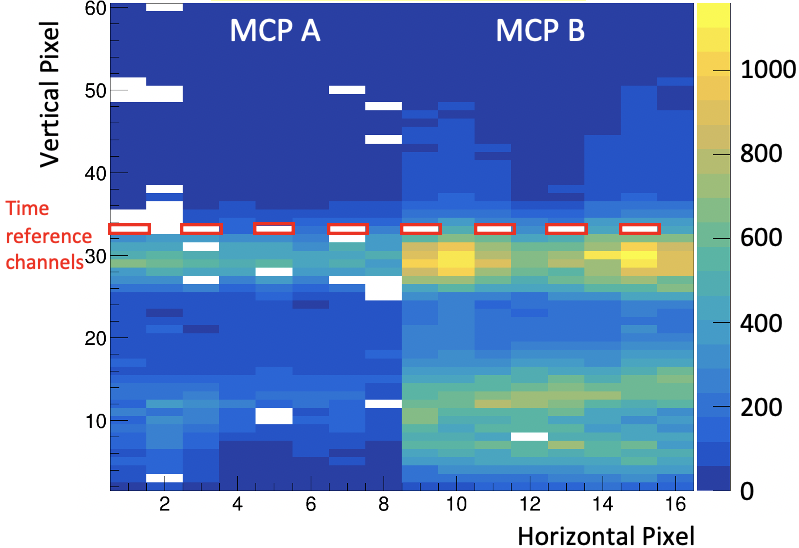} 
\caption{\label{HitmapQE} A hit-map showing photons detected by the two MCP-PMTs. Bands show the different photon paths travelled. Pixels outlined in red are the locations of the time-reference injections. Pixels in white are not read out due to inactive wire bonds.}
\end{center}
\end{figure}

A hit-map of photons detected by the MCP-PMTs is shown in Fig.~\ref{HitmapQE}, where the beam was incident at position 6 on the plate. The pattern shows the projection of the Cherenkov rings on to the MCP-PMT plane, which are observed as curved bands, and which are folded due to reflections of photons from the sides of the plate. The width of the bands is a result of chromatic dispersion. 
The quantum efficiency of both MCP-PMTs was measured independently and was found to be significantly lower for MCP A, which explains the reduced count rate seen in the left half of Fig. \ref{HitmapQE}. 

\begin{figure}[h]
\begin{center}
\includegraphics[width=0.65\linewidth]{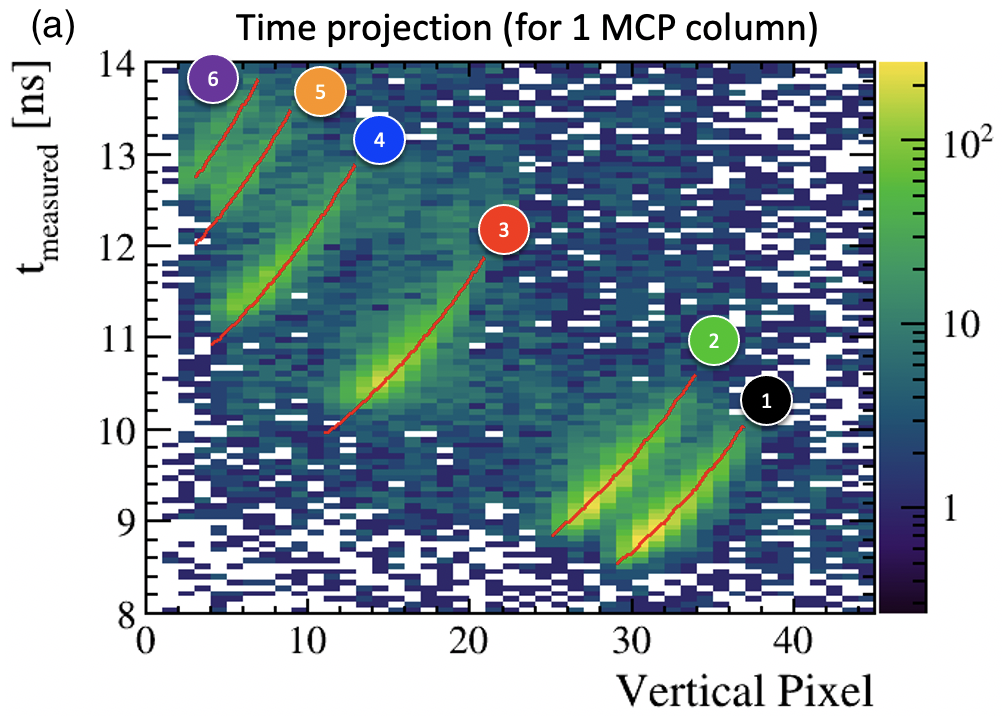} 
\includegraphics[width=0.25\linewidth]{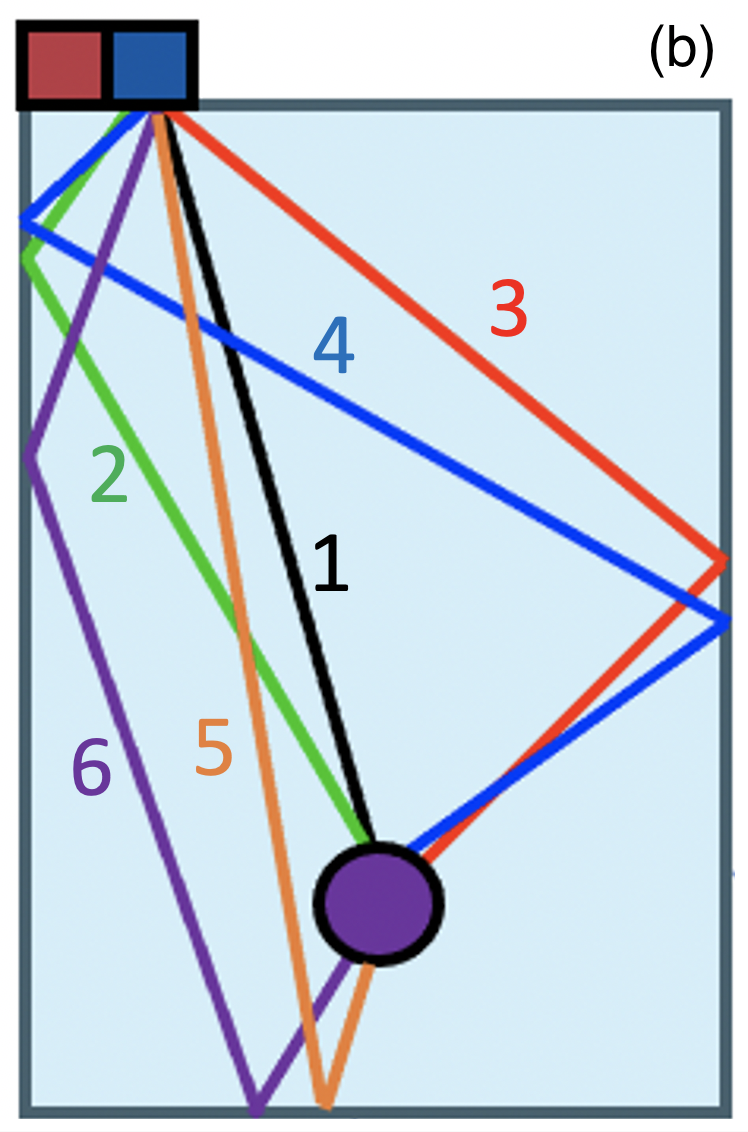}
\caption{\label{TimeplotPaths} (a) The distribution of photon arrival times as a function of the vertical (high granularity) pixel number for photons arriving on a specific column in MCP B. The overlaid lines show the predicted times of arrival. Six distinct bands can be seen which correspond to the photon paths shown in (b).}
\end{center}
\end{figure}

\section{Time Resolution} 

The single-photon time resolution of Proto-TORCH is measured from the spread in photon arrival times for a particular photon path. 
Figure~\ref{TimeplotPaths}a shows a time-projection plot for a single horizontal pixel (low granularity).
The overlaid lines represent the predicted arrival times and match the photon paths shown schematically in Fig.~\ref{TimeplotPaths}b. 
The measured time resolution, $\sigma_{\text{measured}}$, is found by fitting a Crystal Ball function to the distribution of the photon arrival time in each vertical pixel. 
The intrinsic TORCH time resolution, $\sigma_{\text{TORCH}}$, is then given by 
   ${ \sigma_{\text{TORCH}}^2 = \sigma_{\text{measured}}^2 - \sigma_{\text{time ref}}^2 - \sigma_{\text{beam}}^2 }$,
where $\sigma_{\text{time ref}}$ is the resolution due to the time reference input to Proto-TORCH which provides the start time, and $\sigma_{\text{beam}}$ is the contribution due to the finite size of the beam through the radiator plate. 

\begin{figure}[h]
\begin{center}
\begin{minipage}{2.5in}
\includegraphics[width=\linewidth]{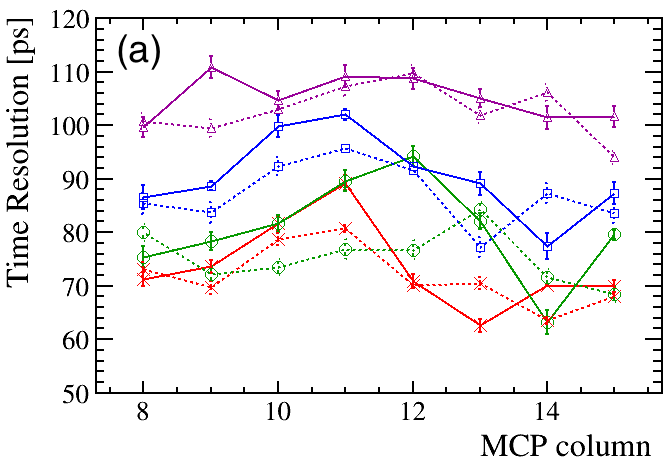} 
\end{minipage}\hspace{2pc}%
\begin{minipage}{2.5in}
\includegraphics[width=\linewidth]{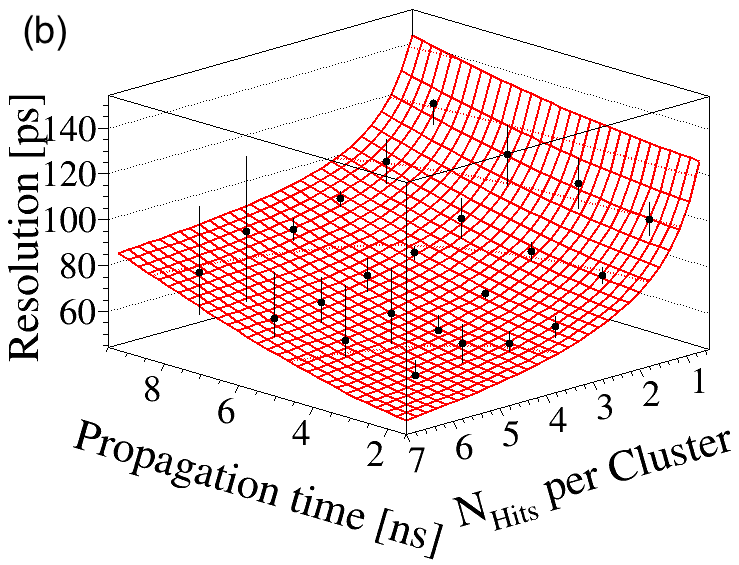}
\end{minipage}
\caption{\label{SingleTR2D} 
(a) 
The TORCH single-photon time resolution as measured across columns in MCP B, and dependent on beam entry positions (1 (red); 3 (green); 4 (blue); 5 (purple)).
The full (dotted) lines indicate results from pions (protons).
(b)
The TORCH single-photon time resolution as a function of photon propagation time within the plate and the number of hits per cluster in the MCP-PMTs. 
}
\end{center}
\end{figure}

Figure~\ref{SingleTR2D}a shows the TORCH time resolution in each horizontal pixel of MCP B for a range of beam entry positions. The 70ps target is attained across pixels for beam entry positions close to the MCP-PMTs. 
It is possible to study the contributions to this resolution by parametrising $\sigma_{\text{TORCH}}^2$ with three terms,
   ${ \sigma_{\text{TORCH}}^2 = \sigma_{\text{MCP}}^2 + \sigma_{\text{prop}}^2(t_p) + \sigma_{\text{RO}}^2(N_{\text{hits}}) }$,
where $\sigma_{\text{MCP}}$ is a constant associated with the resolution of the MCP-PMTs, $\sigma_{\text{prop}}$ is a term dependent on the propagation time, $t_p$, of the photons within the plate, and $\sigma_{\text{RO}}$ is associated with the resolution of the readout electronics and is dependent on the number of hits registered per detected photon in the MCP-PMTs, $N_{\text{hits}}$. 
The measured values from a 2D fit (shown in Fig.~\ref{SingleTR2D}b)  are compared to the target values in Table \ref{tab:ttr2d}. 
Improvements to the measured values are expected with further calibrations, including a measurement of the correlation between the charge deposited and the NINO signal width (needed for improved positioning via charge sharing).

\begin{table}[h]
\caption{\label{tab:ttr2d} Results of the fit to the single-photon time resolution compared to target values from simulation.} 
\begin{center}
\begin{tabular}{l  l l}         
\hline 																					\\[-4pt]
Contribution & Measured value from data (ps) & Target values from simulation (ps) 					\\[5pt] \hline 
$\sigma_{\text{const}}$ & $31.0\pm7.6$ & 33 												\\[5pt]
$\sigma_{\text{prop}}(t_p)$ & $(7.6\pm 0.5)\times t_p [\text{ns}]$ & $(3.75\pm 0.8)\times t_p [\text{ns}]$ 	\\[5pt]
$\sigma_{\text{RO}}(N_{hits})$ & $\frac{95.0\pm6.0}{\sqrt{N_{hits}}}$ & $\frac{60}{\sqrt{N_{hits}}}$ 		\\[5pt] \hline
\end{tabular}
\end{center}
\end{table}

\section{Photon Counting}

The desired track time resolution targets a single-photon resolution of 70ps and the detection of around 30 photons per track. A measurement of the number of detected photon clusters is made for photons that travel directly to the MCP-PMTs (with either no or one near-side reflection) and is compared to Monte Carlo. The optical processes within Proto-TORCH are modelled with GEANT4 \cite{Agostinelli2003} \cite{Allison2006} and custom libraries are used to model the detector and readout responses which include the MCP-PMT quantum and collection efficiencies and the quartz surface roughness. 

\begin{figure}[h]
\begin{center}
\begin{minipage}{6in}
\begin{minipage}{1.95in}
\includegraphics[width=\linewidth]{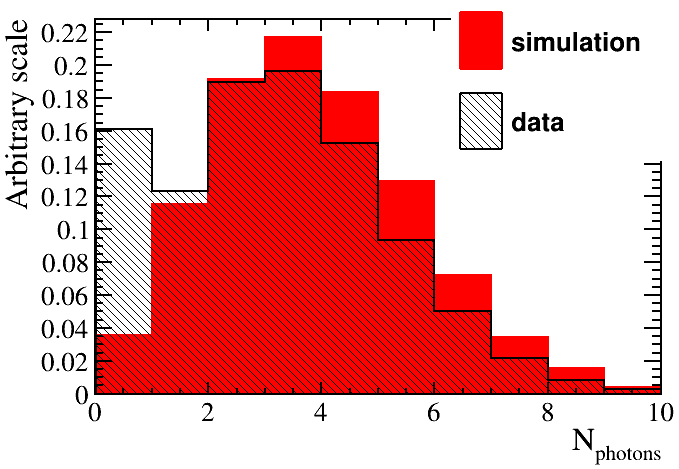} 
\end{minipage}%
\begin{minipage}{1.95in}
\includegraphics[width=\linewidth]{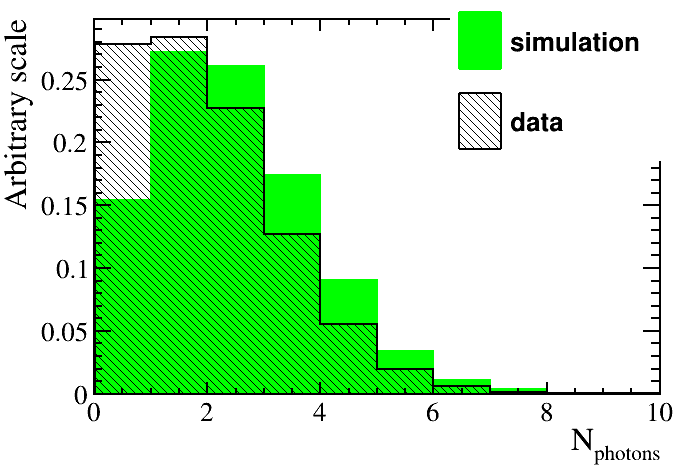}
\end{minipage} %
\begin{minipage}{1.95in}
\includegraphics[width=\linewidth]{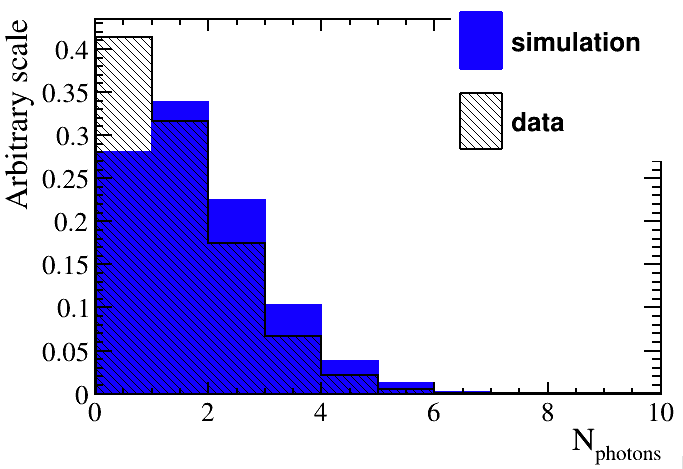} 
\end{minipage}%
\end{minipage} \\
\begin{minipage}{6in}
\caption{\label{PC1345} Number of photons per charged track detected in data (shaded) and simulation (solid) for positions (left-right) 1, 3, 4.}
\end{minipage}
\end{center}
\end{figure}

Figure~\ref{PC1345} shows the distribution of number of photons detected per charged track for three beam entry positions. Good agreement is seen between data and simulation for all positions. There is a small difference in the left-most bin (where there are no photons detected per charged particle) which is attributed to spurious triggers in data-taking. 
The photon yields for all positions are within $90\%$ of Monte Carlo expectations when the zero bin is excluded.
The reduction in photons as a function of distance travelled is a consequence of acceptance effects. 
In a fully instrumented module the photon yield is expected to increase by a factor of 5.5 as there are 11 MCP-PMTs, together with an additional factor due to an increase in the quantum efficiency of the final tubes. 

\section{PID Performance}

\begin{figure}[h]
\begin{center}
\begin{minipage}{2.5in}
\includegraphics[width=\linewidth]{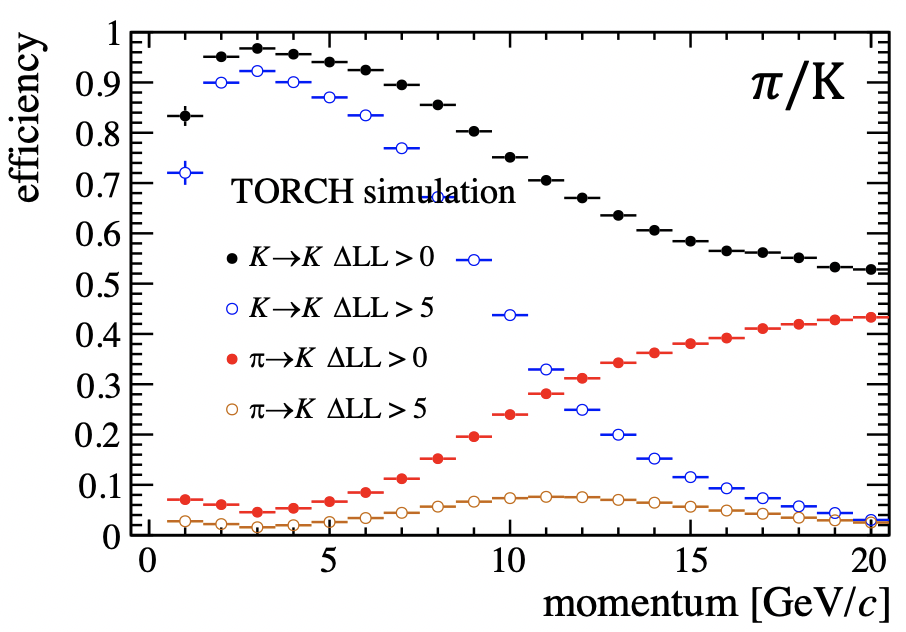} 
\end{minipage}\hspace{2pc}%
\begin{minipage}{2.5in}
\includegraphics[width=\linewidth]{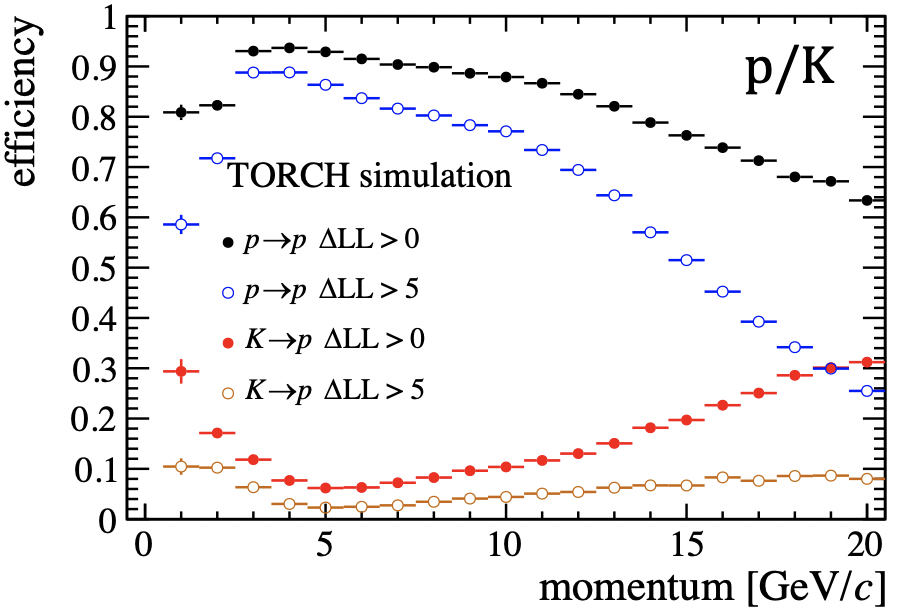}
\end{minipage}
\caption{\label{PID2} PID performance of TORCH from simulation, showing two different cuts on the likelihood differences for (left) pion-kaon separation, (right) kaon-proton separation.}
\end{center}
\end{figure}

The full TORCH detector has been simulated within the LHCb experiment framework \cite{Clemencic2011} in Upgrade II (Run 5) conditions at a luminosity of $\mathcal{L} = 1.4 \times 10^{34}~\text{cm}^{-2} \text{s}^{-1}$. Using a similar method to the LHCb RICH reconstruction process \cite{Adinolfi2013}, the PID is performed by generating the likelihood of each charged track hypothesis by comparing the expected photon patterns. 
Different configurations of pixels and modules are currently under study; here an effective granularity of $128\times32$ is used, where the number of horizontal pixels has been increased compared to the test beam setup to reduce the occupancies. 
The PID performance for pions and kaons and kaons and protons is shown in Fig.~\ref{PID2}; good separation power is achieved up to 10 GeV/c and 20 GeV/c respectively.

\section{Conclusions}
A partially-instrumented half-sized prototype TORCH module has been tested in an 8 GeV/c $\pi$/p beam at CERN. Two custom-designed $53\times53$ mm$^2$ MCP-PMTs were employed, each with an effective granularity of $128\times8$ pixels. 
The single-photon time resolution has been measured and is approaching the design goal of 70 ps. Improvements are expected with further calibrations of the readout electronics system. 
The measured photon yields are close to expectations.
Simulations of TORCH in the LHCb Upgrade II conditions show that efficient pion-kaon and kaon-proton discrimination can be achieved up to 10 GeV/c and 20 GeV/c respectively.  

\section*{Acknowledgments}
The support is acknowledged of the Science and Technology Research Council, UK, grant number ST/P002692/1, and of the European Research Council through an FP7 Advanced Grant (ERC-2011-AdG 299175-TORCH). 

\printbibliography
\end{document}